\documentclass[journal,a4paper]{IEEEtran}
\usepackage{multicol}
\usepackage{graphicx, xcolor}
\usepackage{cite}
\usepackage{amssymb, amsthm}
\usepackage{mathtools}  
\usepackage{multirow}
\usepackage{scalerel}
\usepackage{wrapfig}
\usepackage{caption}
\usepackage{subcaption}
\usepackage{empheq}
\usepackage{booktabs}
\usepackage{graphics}
\usepackage{atbegshi}
\AtBeginDocument{\AtBeginShipoutNext{\AtBeginShipoutDiscard}}

\makeatletter
\newcommand*{\rom}[1]{\expandafter\@slowromancap\romannumeral #1@}
\makeatother
\newtagform{normalsize}[\normalsize]{\normalsize(}{\normalsize)}

\begin{document}

\title{Event-Triggered Non-Linear Control of Offshore MMC Grids for Asymmetrical AC Faults}

\author{Naajein~Cherat,~Vaibhav~Nougain, Milovan Majstorović, Peter Palensky, \textit{Senior Member, IEEE}, and~Aleksandra Leki\'{c}, \textit{Senior Member, IEEE}}

\thanks{N. Cherat, V. Nougain, P. Palensky, and A. Leki\'{c} are with the Department of Electrical Sustainable Energy, Delft University of Technology, Delft, The Netherlands. (e-mail: N.Cherat@student.tudelft.nl, \{V.Nougain, P.Palensky, A.Lekic\}@tudelft.nl).

M. Majstorović is with the University of Belgrade, Belgrade, Serbia. (e-mail: majstorovic@etf.bg.ac.rs)

This project has received funding from the European Union’s HORIZON-WIDERA-2021-ACCESS-03 under grant agreement No. 101079200.
}

\IEEEoverridecommandlockouts
\IEEEpubid{\makebox[\columnwidth]{979-8-3503-9042-1/24/$\$$31.00
\copyright2024 European Union \hfill} \hspace{\columnsep}\makebox[\columnwidth]{ }}

\maketitle
\begin{abstract}
Fault ride-through capability studies of MMC-HVDC connected wind power plants have focused primarily on the DC link and onshore AC grid faults. Offshore AC faults, mainly asymmetrical faults have not gained much attention in the literature despite being included in the future development at national levels in the ENTSO-E HVDC code. The proposed work gives an event-triggered control to stabilize the system once the offshore AC fault has occurred, identified, and isolated. Different types of control actions such as proportional-integral (PI) controller and super-twisted sliding mode control (STSMC) are used to smoothly transition the post-fault system to a new steady state operating point by suppressing the negative sequence control. Initially, the effect of a negative sequence current control scheme on the transient behavior of the power system with a PI controller is discussed in this paper. Further, a non-linear control strategy (STSMC) is proposed which gives quicker convergence of the system post-fault in comparison to PI control action. These post-fault control operations are only triggered in the presence of a fault in the system, i.e., they are event-triggered. The validity of the proposed strategy is demonstrated by simulation on a $\pm$525 kV, three-terminal meshed MMC-HVDC system model in Real Time Digital Simulator (RTDS).
\end{abstract}


%

\section{Introduction}
\IEEEPARstart {O}{ffshore} wind farms are experiencing rapid growth as a sustainable energy solution, attributed to their advantages of reduced wind variability and space constraints \cite{ref12}. High voltage direct current (HVDC) is a proven technology for the grid integration of offshore wind farms.
Modular multilevel converters (MMC) are used in HVDC transmission systems due to their distinctive features such as modular structure, high reliability, effective redundancy, simple fault identification, and clearance \cite{ref5}. Owing to its grid-forming ability and control capability, the MMC-HVDC-based power system is considered a cost-effective option for the integration of offshore wind farms \cite{ref4}. 

One of the critical contingencies affecting the stability of MMC-based HVDC systems is the short circuit fault \cite{roose2021}. This can result in power system instability. An effective fault ride-through strategy is critical to avoid HVDC converter station disconnection from the AC grid. The literature for research on short-circuit faults in HVDC systems has primarily focused on the DC link and onshore AC grid faults \cite{ref6}, neglecting the offshore AC fault contingency. Asymmetrical AC faults (single-phase-to-ground, phase-to-phase, and phase-to-phase-to-ground) represent the most probable fault contingencies in the AC power system \cite{ref7}. On the occurrence of an asymmetrical fault, large negative sequence currents will be generated in the system. Due to the over-current limitations imposed by the converters, the control strategy of negative sequence current under asymmetric AC faults is of great significance \cite{ref7}. 

During an unbalanced onshore grid fault, the impact of negative sequence current control schemes on an onshore AC transmission line is analyzed in \cite{ref8}. Two scenarios are considered, the suppression of negative sequence current and injection of negative sequence current proportionally to the negative sequence voltage. It is shown that injecting negative sequence current enhances fault detection capabilities and improves the performance of protection schemes in comparison to suppression of negative sequence current. 

Work \cite{ref1} proposes a control strategy that utilizes negative sequence voltages to facilitate controlled injection of negative sequence currents during offshore asymmetric AC faults. It is shown that by actively managing negative sequence currents and voltages, the strategy ensures a controlled level of fault current and prevents overvoltage conditions in healthy phases post-fault. This approach not only enhances the stability and reliability of the offshore wind power transmission system but also minimizes the risk of protection mal-operation. An enhancement to the control strategy proposed in \cite{ref1} is done in \cite{ref4}. The per unit value of the MMC valve-side voltage positive sequence component is used as the droop coefficient, aiming to reduce over-adjustment issues in AC voltage during fault periods. The over-adjustment of AC voltage can lead to a significant reduction in the active power transmission of the HVDC system in case of a large voltage drop.

Paper \cite{ref2} proposes a high-performance fault ride-through method for an MMC-integrated offshore wind farm system, showcasing responses to symmetric and asymmetric faults. A method for continuous power transmission under faults and effective suppression of overcurrent and modulation is explained in \cite{ref2}. Negative sequence current coordinated control strategy for severe asymmetric offshore AC faults is discussed in \cite{ref7}. 
When there is a severe asymmetric AC fault, the control strategy ensures that the negative sequence current of the offshore MMC is suppressed to zero, while the GSC (grid-side converters) cooperates to reduce the positive sequence current. The strategy involves limiting the positive sequence current of the GSC and reducing the positive sequence voltage of the offshore MMC to reserve margins for the modulation of the negative sequence voltage.

This paper focuses on an event-triggered suppression of negative sequence current to improve the overall transient behavior of the power system during an unbalanced offshore grid fault, restoring system stability after the fault is cleared. Linear (PI) and non-linear (STSMC) controller are tested for their effectiveness in restoring the system. Since non-linear controllers have a faster convergence compared to their linear PI counterparts, the proposed non-linear controller is well suited to counter the non-linear dynamics of MMC converters. The proposed control strategy is tested on a $\pm$525 kV three-terminal meshed MMC-HVDC system modeled in Real Time Digital Simulator (RTDS). The rest of the paper is organized as follows. Section \rom{2} outlines the system description and the main controllers used. Section \rom{3} explains the implementation of negative sequence control. Section \rom{4} validates the proposed controller for various asymmetrical AC faults. Finally, Section \rom{5} provides the conclusion and scope for future work.

\section{SYSTEM DESCRIPTION}
\begin{figure}[!thb] 
		\centering
	\includegraphics[width=\columnwidth]{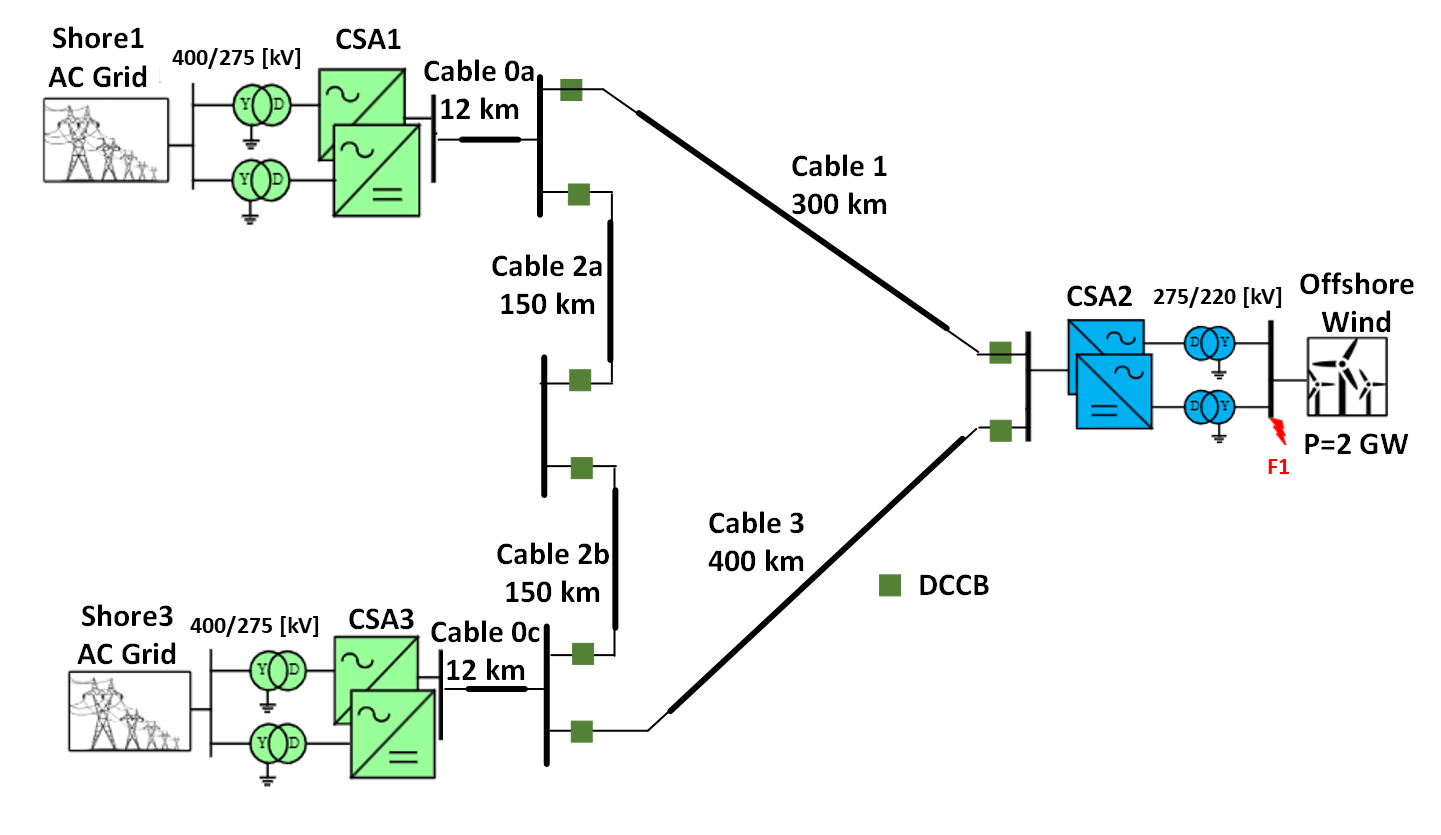}
	\caption{Configuration of the $\pm$525 kV, 2GW, three-terminal MMC-HVDC system}
	\label{fig:system}
\end{figure}

Figure \ref{fig:system} illustrates the configuration of the three-terminal meshed MMC-HVDC system used in this study. The system consists of an offshore wind plant, an offshore converter, and two onshore converters. The rated power of the MMC-HVDC system is 2GW and the rated voltage is $\pm$525 kV. A bipolar configuration is used for HVDC in this study. Converters CSA1 and CSA3 are connected to strong AC grids. The offshore wind farms are connected to the DC grid through the converter CSA2. A cable of 300 km in length and 400 km in length connect the CSA2 with CSA1 and CSA3 respectively. The onshore converters CSA1 and CSA3 are also connected via cables. 
\vspace*{-2mm}

Two-level control typical for the Voltage Source converter (VSC) is used for the MMC converters. The upper level of control generates the reference voltage wave. The lower level control provides valve firing signals and ensures that cell capacitor voltage in each submodule (SM) remains constant according to a predetermined value. The reference is generated by using vector control philosophy where the 3-$\phi$ ABC system quantities are transformed into DQ system quantities by Park's transformation. This is done to have a flexible and independent control of active and reactive power using simple DC parameters (in the DQ system) of their equivalent 3-$\phi$ parameters (in the ABC system). The upper-level control has a decoupled inner current controller whose references are based on user-defined MMC operating mode.

\begin{figure}[!thb] 
		\centering
	\includegraphics[width=0.9\columnwidth]{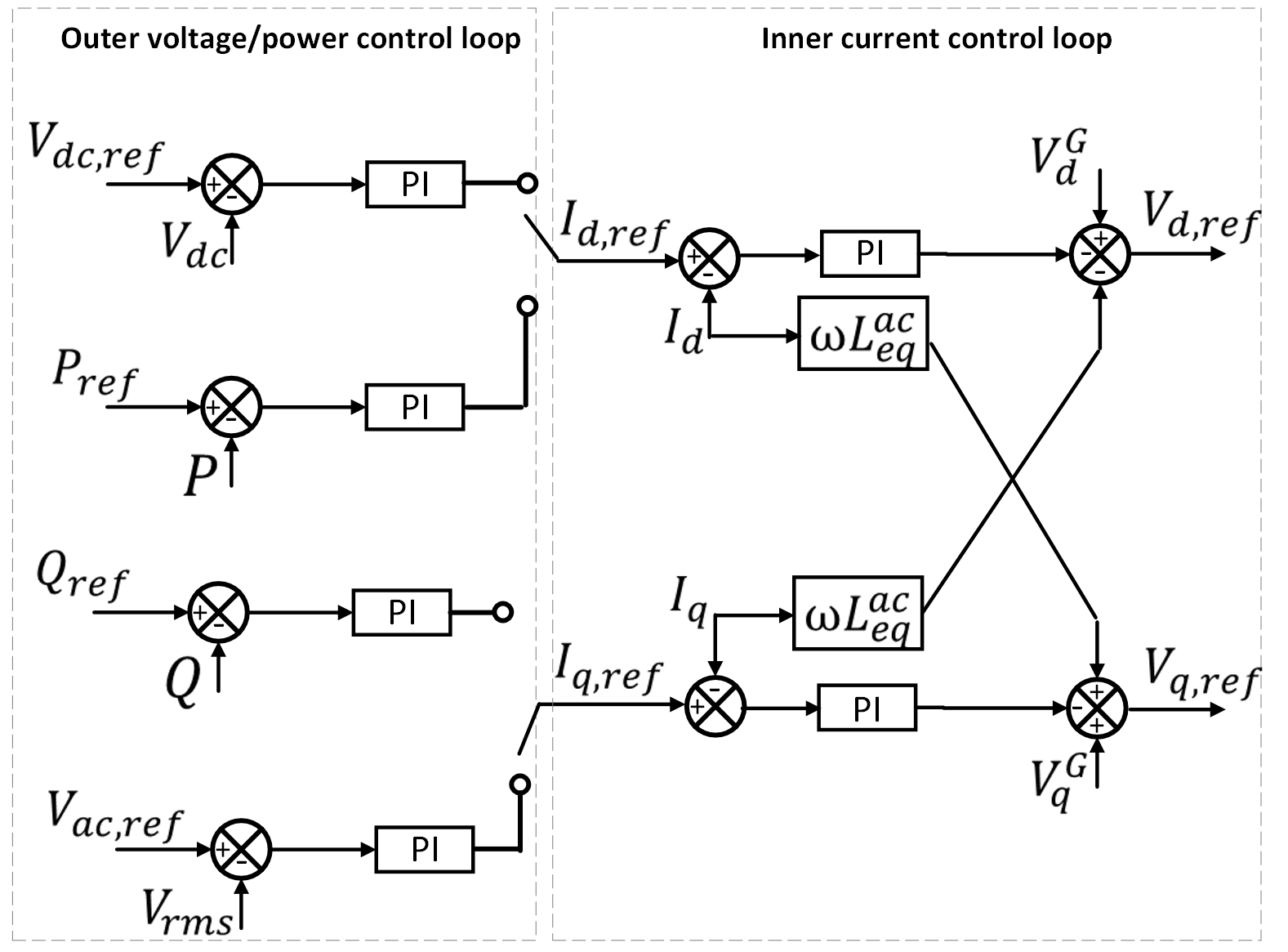}
	\caption{MMC control for grid-following converters}
	\label{fig:GFL}
\end{figure}

Onshore converter, CSA1 acts as the slack converter controlling the DC voltage of the whole system. The control mode used is $V_{dc}/V_{ac}$, where the d-axis current reference $i_{dref}$ is controlled by the DC voltage loop and q-axis current reference $i_{qref}$ is controlled by AC voltage loop. Onshore converter, CSA3 uses control mode $P_{ac}/Q_{ac}$ where the d-axis current reference, $i_{dref}$ is controlled by the active power loop and q-axis current reference, $i_{qref}$ is controlled by the reactive power loop. The control loops for CSA1 and CSA3 come under the class of grid-following (GFL) control as shown in Fig. \ref{fig:GFL}. GFL converter requires a dedicated unit to identify the grid voltage angle and calculate the proper phase shift. This is done using phase-locked loops (PLLs).
\vspace*{-2mm}

Lower-level control keeps the capacitor voltages across all submodules (SMs) within an acceptable range. This is achieved by selectively switching SMs based on the direction of arm currents. However, this regulation of capacitor voltages results in circulating currents among the three-phase units. This current does not influence the currents in the AC and DC sides, but they distort the
arm current and increase the rated current of the submodules. A control loop to suppress the circulating current is included to reduce the effects of the circulating current. 


The MMC model used in the present study assumes that the capacitor voltages of each submodule (SM) are internally balanced. Hence, there's no need to specify the specific SMs for insertion; only the total number of SMs needs to be specified. The control input is thus simplified to an overall deblock integer signal and the number of SMs to be inserted.

\begin{figure}[!thb] 
		\centering
\includegraphics[width=0.5\columnwidth]{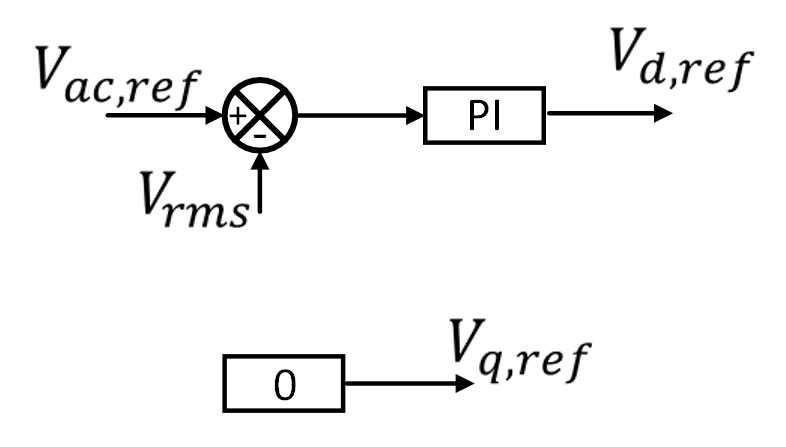}
	\caption{MMC control for offshore converter}
	\label{fig:GFM}
\end{figure}

 The offshore converter, CSA2 uses a grid-forming (GFM) control. GFM converters are responsible for establishing and regulating grid voltages at the point of common coupling (PCC), especially in islanded operation mode. The control objective is to stabilize grid frequency and regulate voltage amplitude. GFM converters can self-synchronize to the grid without the need for a dedicated unit. An overview of the different control algorithms for GFM converters is provided in \cite{ref10}. Out of the variety of voltage control schemes developed, the simplest approach is to directly feed the voltage magnitude and phase angle which are generated from outer control loops (Fig. \ref{fig:GFM}) to the modulation . 
 Lack of current controllability may cause overcurrent tripping of the converter during grid faults. During the fault, control loop is switched to a dual loop consisting of an inner current loop which will be explained further in section \rom{3}.
\vspace*{-2mm}

In this study, a Type 4 wind turbine is used, connecting the stator of the permanent magnet synchronous machine (PMSM) to the grid via two full-scale back-to-back converters. VSCs linked to the grid regulate DC voltage and reactive power, while those connected to the PMSM reduce reactive power and enhance generator efficiency.
\section{IMPLEMENTATION OF NEGATIVE SEQUENCE CONTROL}
In the event of unbalanced faults, conventional generation units typically act as voltage sources in the positive sequence circuit, leading to high fault currents. However, power converter-interfaced generation units function as controlled current sources in both the positive and negative sequence circuits, while the zero sequence circuit remains mostly open due to transformer configuration \cite{ref8}. When there is an asymmetrical fault, large negative sequence currents are generated in the system. A common strategy during unbalanced fault conditions is suppressing the negative sequence current to prevent the switching valves from uncontrolled fault currents.

\begin{figure}[!thb] 
	\centering	\includegraphics[width=0.9\columnwidth]{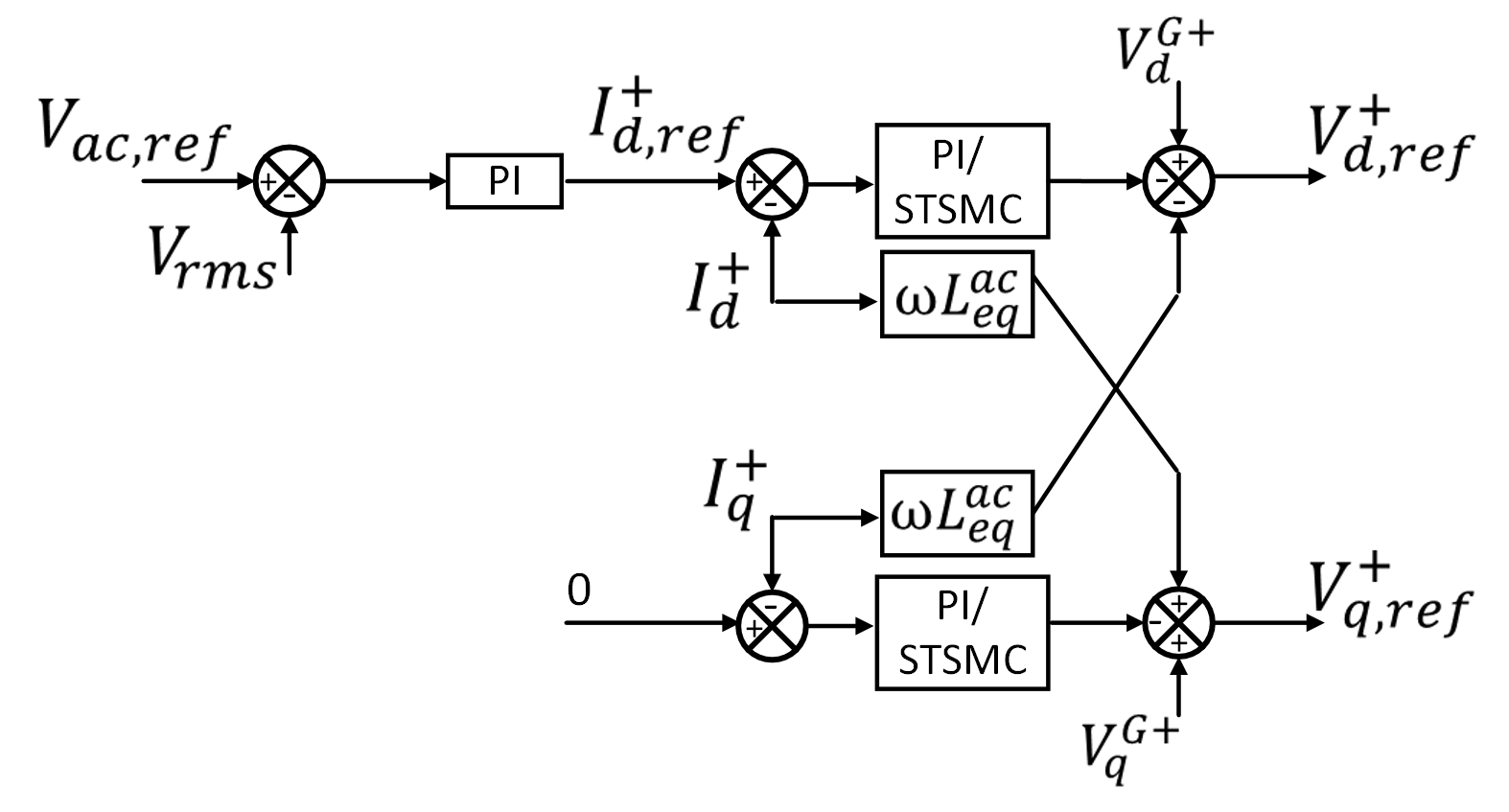}
	\caption{MMC offshore converter control - Positive sequence}
	\label{fig:pos}
	\includegraphics[width=0.6\columnwidth]{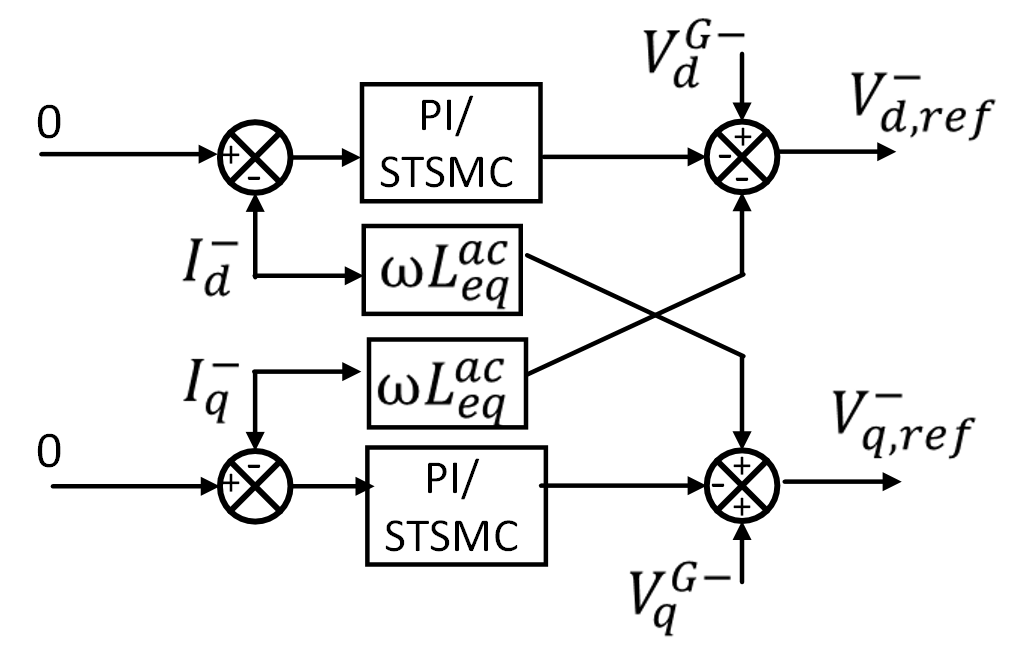}
	\caption{MMC offshore converter control - Negative sequence}
	\label{fig:neg}
\end{figure}

When a fault is detected, the sequence components are used to control the offshore converter CSA2. An event-triggered control strategy is activated to regulate the sequence components of voltage and current signals during a fault. The sudden increase in negative sequence current serves as a fault detection mechanism, triggering the proposed control strategy. 
The presence of a fault is detected by comparing the negative sequence current across two consecutive time steps; if the difference exceeds a set limit, a fault is indicated. The value considered in this study is 0.45 p.u. The positive and negative sequence control for the offshore MMC is given in Fig. \ref{fig:pos}-\ref{fig:neg} respectively.

Conventional PI controllers are implemented in all control loops of the model. Numerous studies have underscored the advantages of employing non-linear controllers over conventional PI controllers to enhance power system performance. The authors in \cite{ref13, 9968871} explain the different non-linear controllers such as Model Predictive Control (MPC), Back-Stepping Control (BSC), and Sliding Mode Control (SMC). Out of the three, SMC has the benefit of high robustness against system uncertainties and fast transient response. However, the chattering problem of SMC can lead to system oscillations \cite{ref14}. A second-order SMC, super twisting sliding mode controller (STSMC), has become popular recently, which not only preserves the robustness of the classical SMC but also overcomes the chattering issue.

In the SMC controller, a sliding surface is defined that guides the system states towards desired values. This sliding surface is typically based on tracking error, with the goal of minimizing this error to zero. When the states reach this surface, the controller uses a sign function to maintain the condition, which can lead to chattering phenomena due to the discontinuous nature of the sign function. Higher-order SMC such as STMSCs, are used to address the chattering issues and improve the performance of the controller. 

The sliding surface $\dot{S}$ for STSMC is proposed as outlined in Eq. \ref{eq:STSMC} \cite{ref13}. The value of $x$ is determined using the error signal obtained from comparing the measured actual value $y(t)$ to the reference value $y^*(t)$ as given in Eq. \ref{eq:STSMCx}. Since both the terms of the sliding surface are continuous, the chattering is reduced. 
\begin{equation}
    \dot{S}(y) = -\alpha\sqrt(x)\operatorname{sgn}(x)-\beta\int(\operatorname{sgn}(x)), \alpha,\beta > 0
    \label{eq:STSMC}
\end{equation}
\begin{equation}
e = y^*(t) - y(t), 
x = K_ie+K_Pe ,
\label{eq:STSMCx}
\end{equation}
\noindent where $\alpha=1.5\sqrt{H}$ and $\beta = 1.1 H$. \textit{H} is the upper bound of system disturbance. For this study, the value of H is set to 10. 
\newline
The traditional PI controller within the inner current control loop has been substituted with STSMC for regulating sequence components in the offshore MMC, CSA2. The remaining control loops, including outer control loops, circulating current suppression control, and onshore converter control loops, continue to utilize state-of-the-art PI controllers. 
\section{SIMULATION STUDIES}

{\renewcommand{\arraystretch}{1}
\begin{table}
\centering
\caption{Control mode and parameters of MMC converters}
\label{tab:sys_para}
\resizebox{\linewidth}{!}{%
\begin{tabular}{llll} 
\hline
\multirow{2}{*}{\textbf{Parameter}} & \multicolumn{3}{c}{\textbf{Converters}} \\
 & \textbf{CSA 1} & \textbf{CSA2} & \textbf{CSA3} \\ 
 \\
\hline
\textbf{Control mode} & $V_{dc}/V_{ac}$ & $V_{ac}/f$ & $P_{ac}/Q_{ac}$ \\
\textbf{Rated active power [MW]} & 2000 & 2000 & 2000 \\
\textbf{AC Grid Voltage [kV]} & 400 & 220 & 400 \\ 
\hline
\textbf{DC link Voltage [kV]} &\multicolumn{3}{c}{$\pm$525}\\
\textbf{Number of Submodules per arm} &\multicolumn{3}{c}{200}\\
\textbf{MMC arm inductance [mH]} &\multicolumn{3}{c}{39.7}\\
\textbf{MMC arm capacitance [$\mu$F]} &\multicolumn{3}{c}{15000}\\
\textbf{Transformer leakage reactance [pu]} &\multicolumn{3}{c}{0.18}\\
\textbf{AC converter bus voltage [kV]} &\multicolumn{3}{c}{275}\\
\hline
\end{tabular}}
\end{table}
}

This section presents an in-depth analysis of the transient response of a 3-terminal meshed MMC-HVDC system under various asymmetrical faults (L-G, L-L, and L-L-G). Table \ref{tab:sys_para} provides details of the control mode and parameters of the MMC converters. 
As illustrated in Fig. \ref{fig:system}, the fault occurs on the line connecting the offshore wind plant and the offshore MMC (CSA2). It is assumed that the AC fault will clear on its own after 300 milliseconds.  A fault resistance of 0.001 $\Omega$ is considered. 

\begin{figure}[!thb] 
		\centering
	\includegraphics[width=\columnwidth]{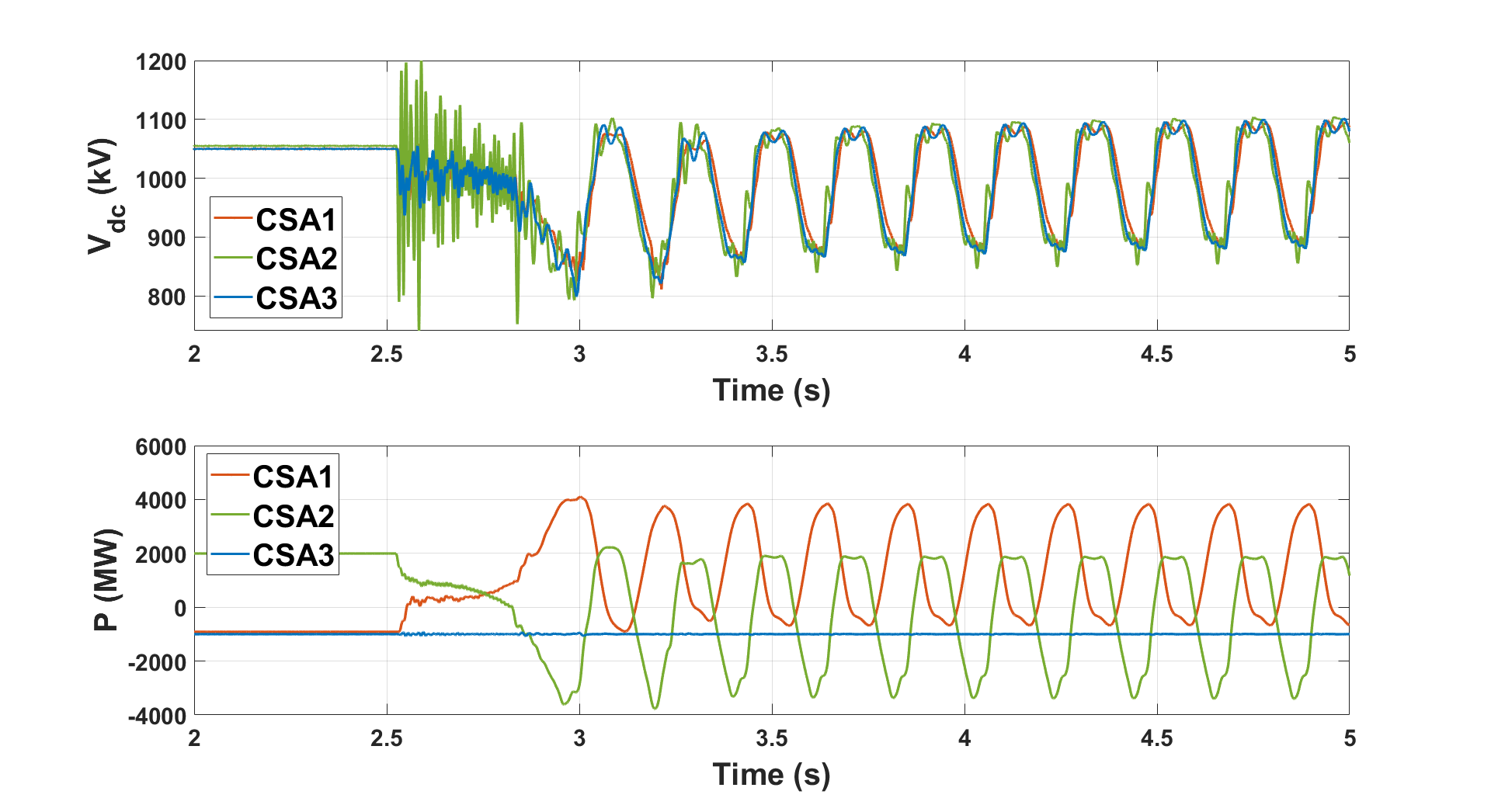}
	\caption{DC link voltage and active power plots for different MMC's}
	\label{fig:default}
\end{figure}

Initially, the fault is introduced to the system without applying the proposed sequence current control. The most severe fault scenario, L-L-G, is chosen to evaluate the system's response. The DC link voltage and active power signal for the different converters are given in Fig. \ref{fig:default}. It is observed that once the fault is cleared, the system remains unstable and the voltage magnitude and phase angle control is not able to stabilize the system. The active power control of 1 GW of the onshore converter CSA3 makes the power signal for CSA3 remain constant.

\begin{figure}[!thb] 
		\centering
	\includegraphics[width=\columnwidth]{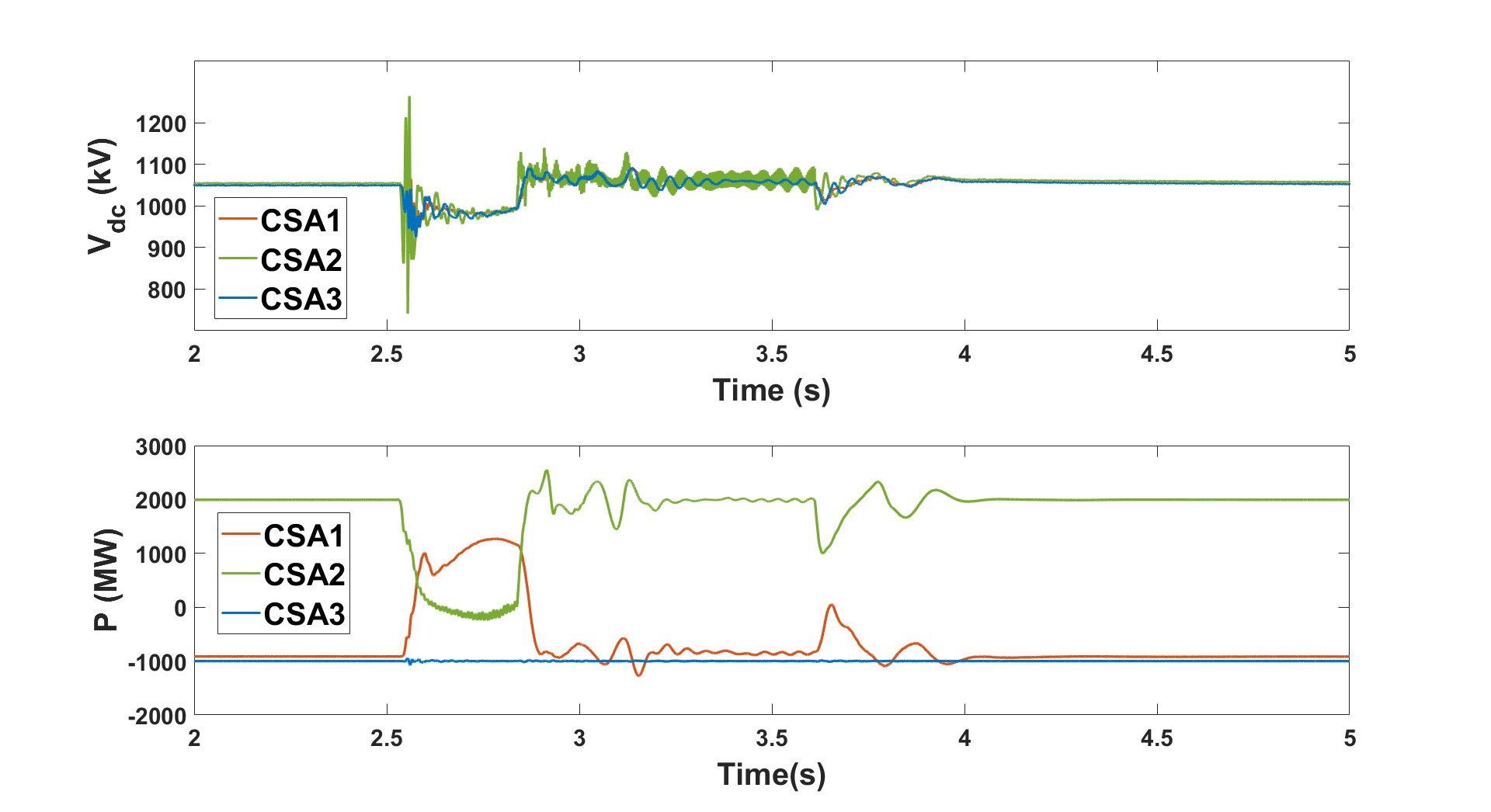}
	\caption{DC link voltage and active power plots for different MMC's during L-L-G fault using PI controller}
	\label{fig:LLG}

    \includegraphics[width=\columnwidth]{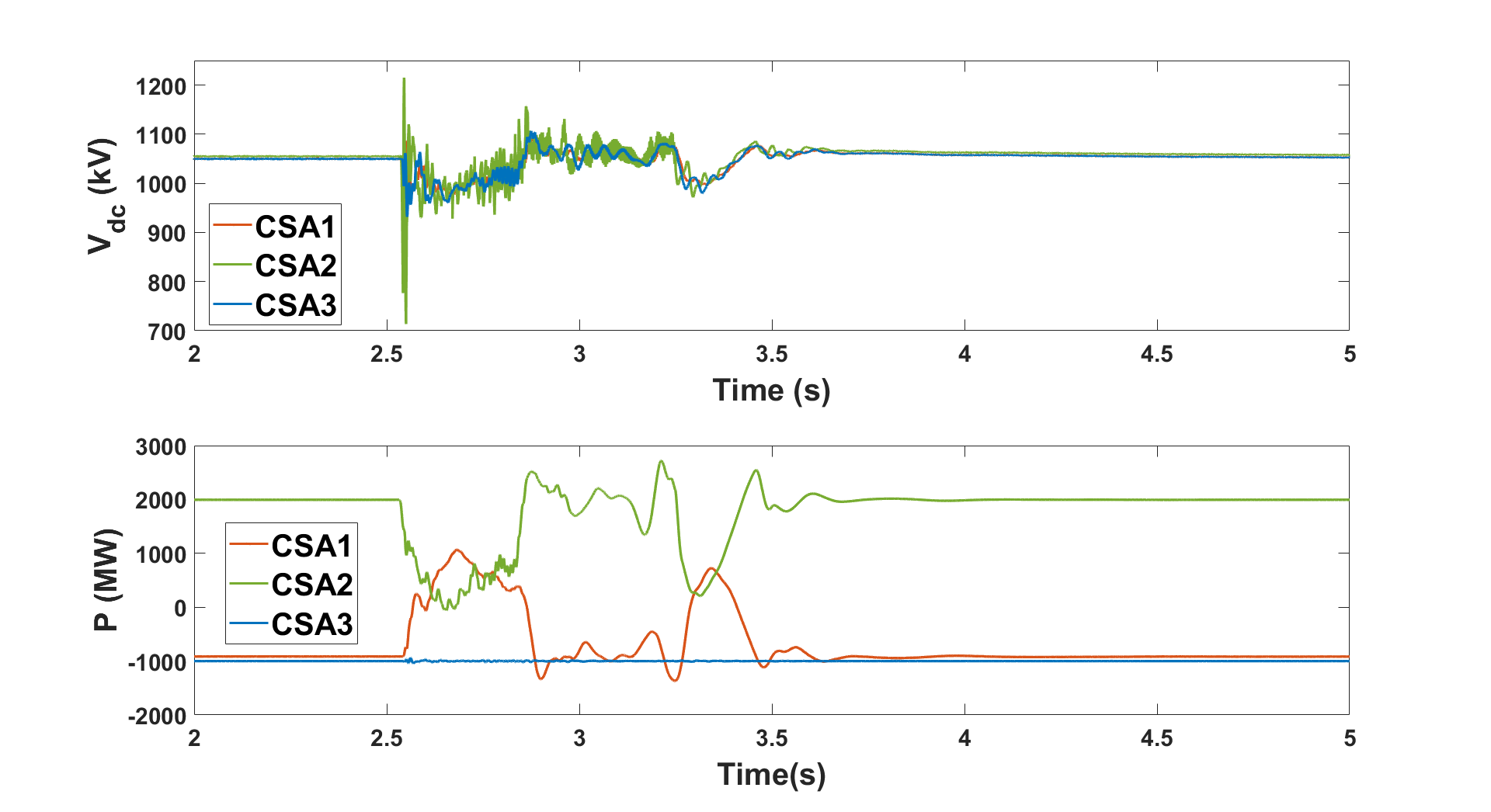}
	\caption{DC link voltage and active power plots for different MMC's during L-L fault using PI controller}
	\label{fig:LL}

	\includegraphics[width=\columnwidth]{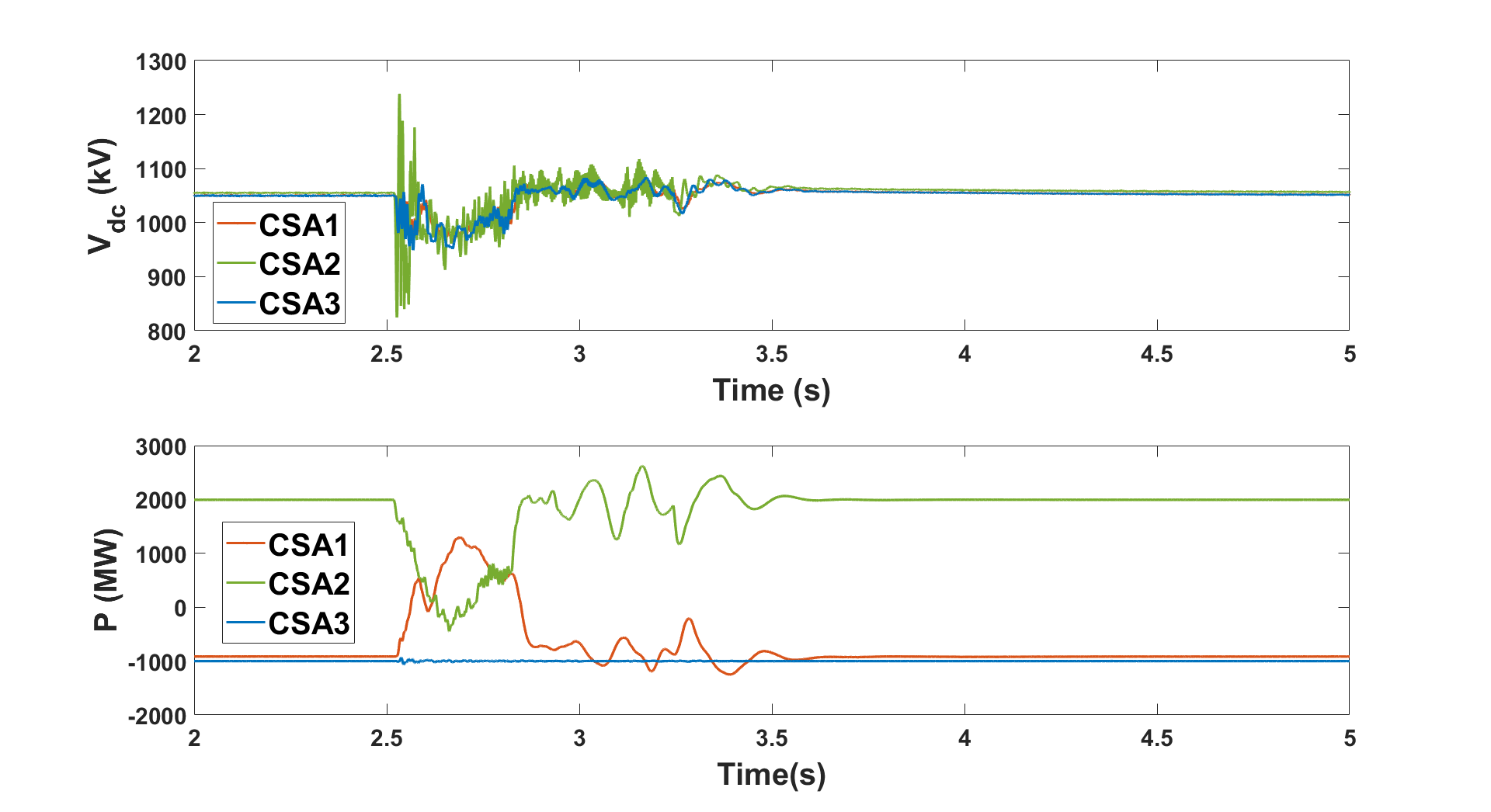}
	\caption{DC link voltage and active power plots for different MMC's during L-G fault using PI controller}
	\label{fig:LG}
\end{figure}

Subsequently, the sequence current control scheme is activated upon fault occurrence. The DC link voltage and active power signals for different faults (L-G, L-L, and L-L-G) using the conventional PI controller are given in Fig. \ref{fig:LLG}, \ref{fig:LL} and \ref{fig:LG}. It is evident that with the introduction of sequence control, the system reverts back to a stable state post fault clearance. Due to its severity, the L-L-G fault necessitates a longer period for the system to stabilize compared to less severe faults. 

\begin{figure}[!thb] 
		\centering
	\includegraphics[width=\columnwidth]{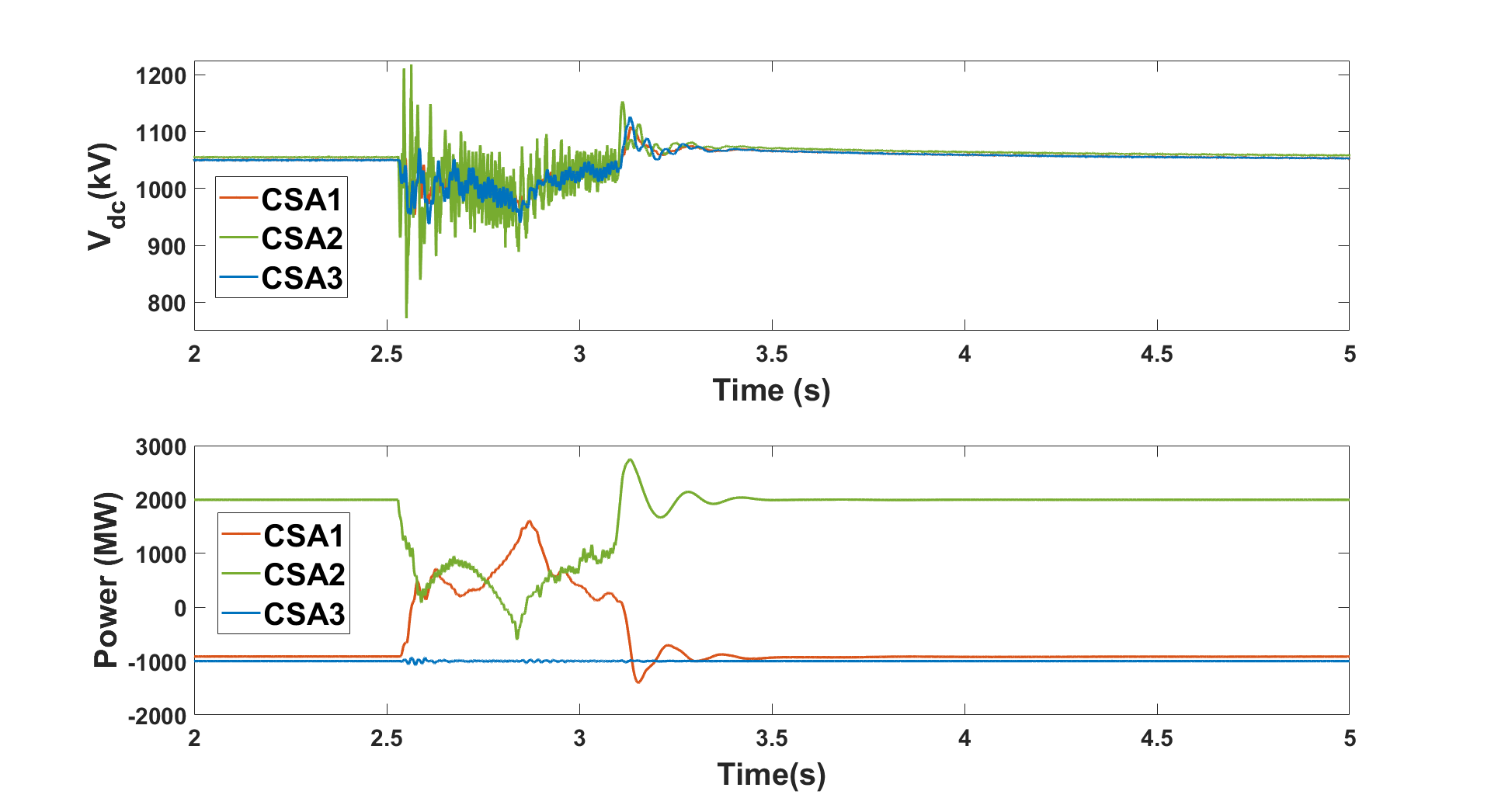}
	\caption{DC link voltage and active power plots for different MMC's during L-L-G fault using STSMC controller}
	\label{fig:LLG_STSMC}
 
		\centering
  	\includegraphics[width=\columnwidth]{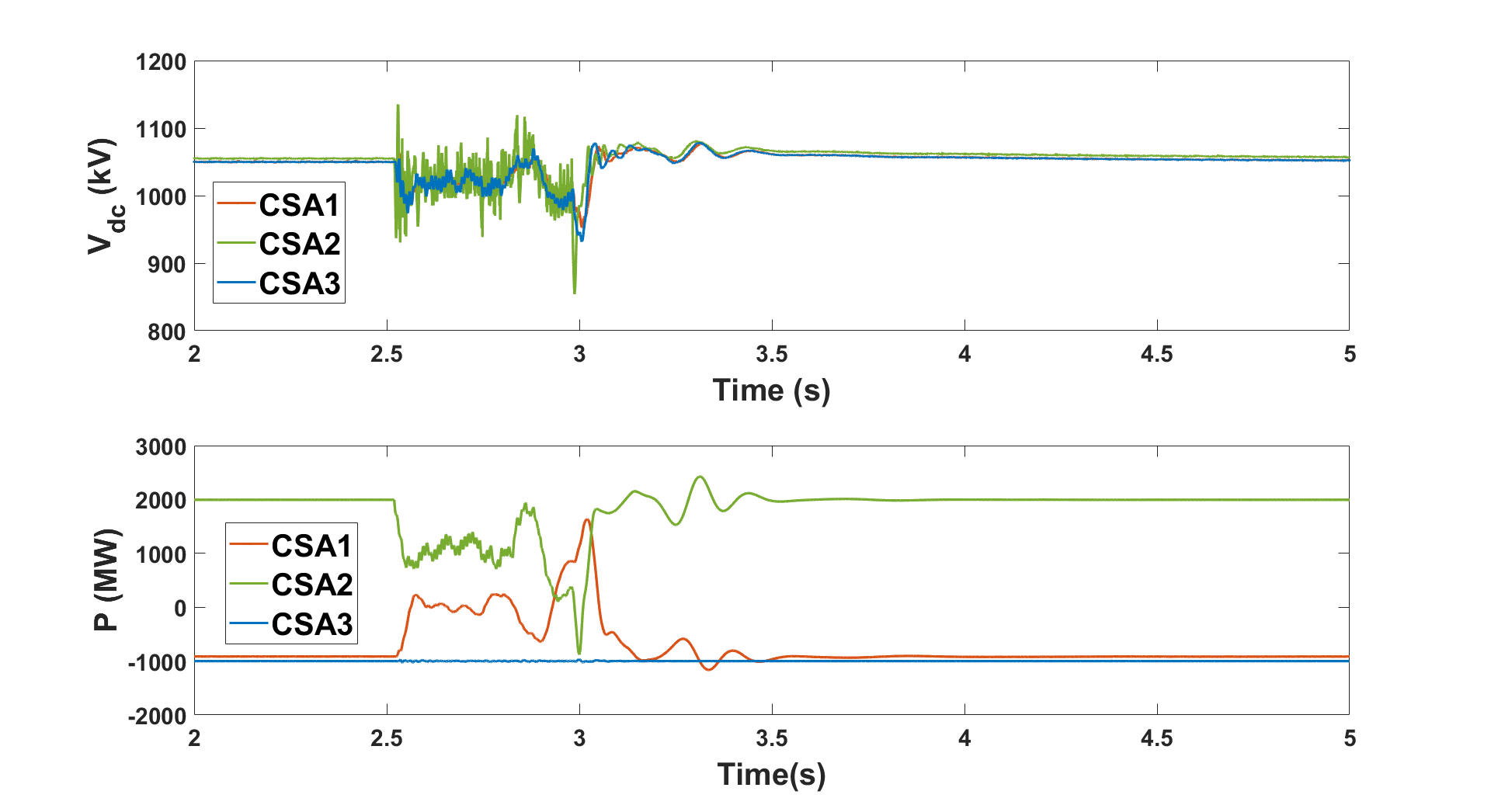}
	\caption{DC link voltage and active power plots for different MMC's during L-L fault using STSMC controller}
	\label{fig:LL_STSMC}
 
		\centering
  	\includegraphics[width=\columnwidth]{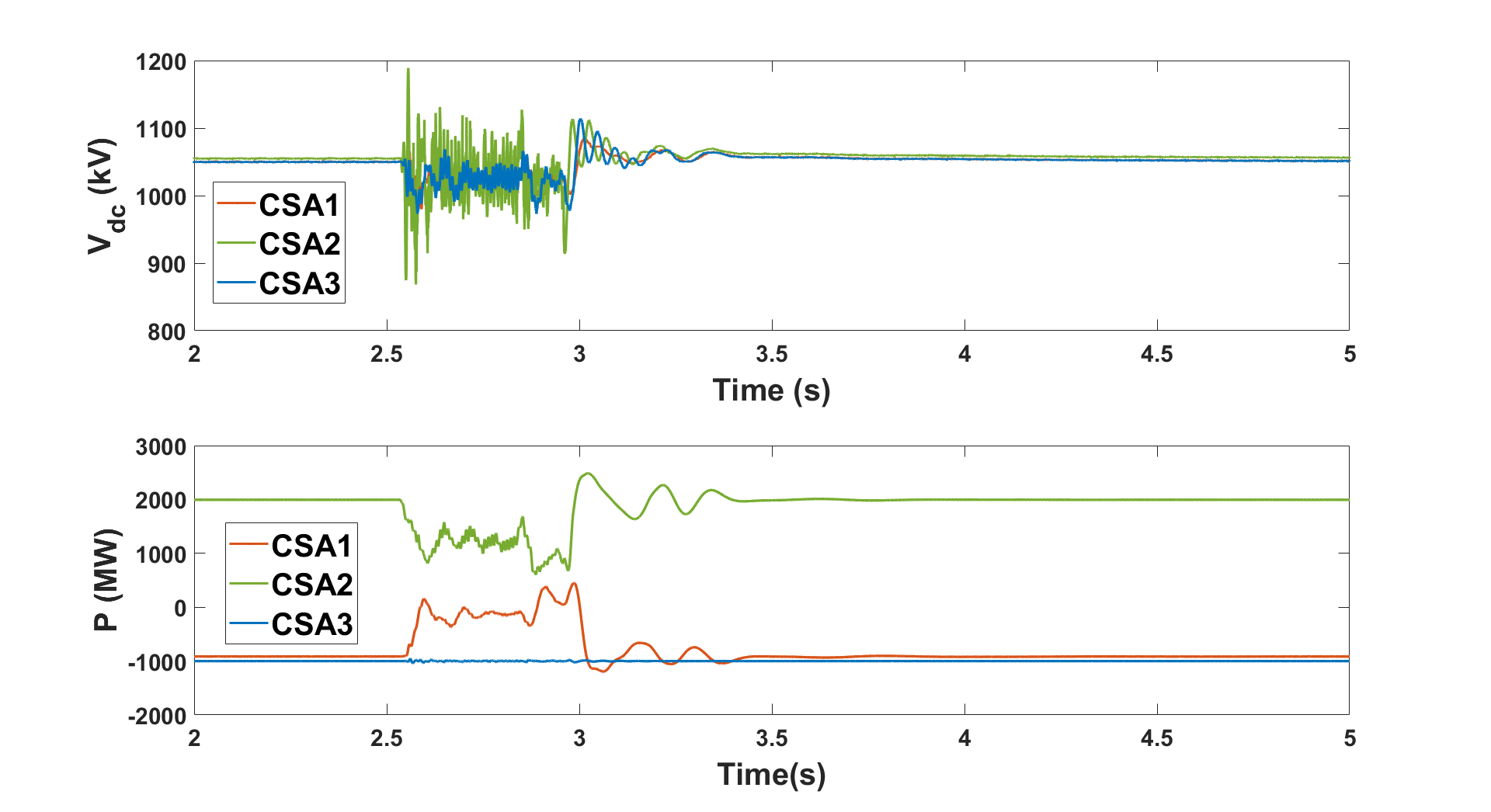}
	\caption{DC link voltage and active power plots for different MMC's during L-G fault using STSMC controller}
	\label{fig:LG_STSMC}
 
 		\centering
   	\includegraphics[width=\columnwidth]{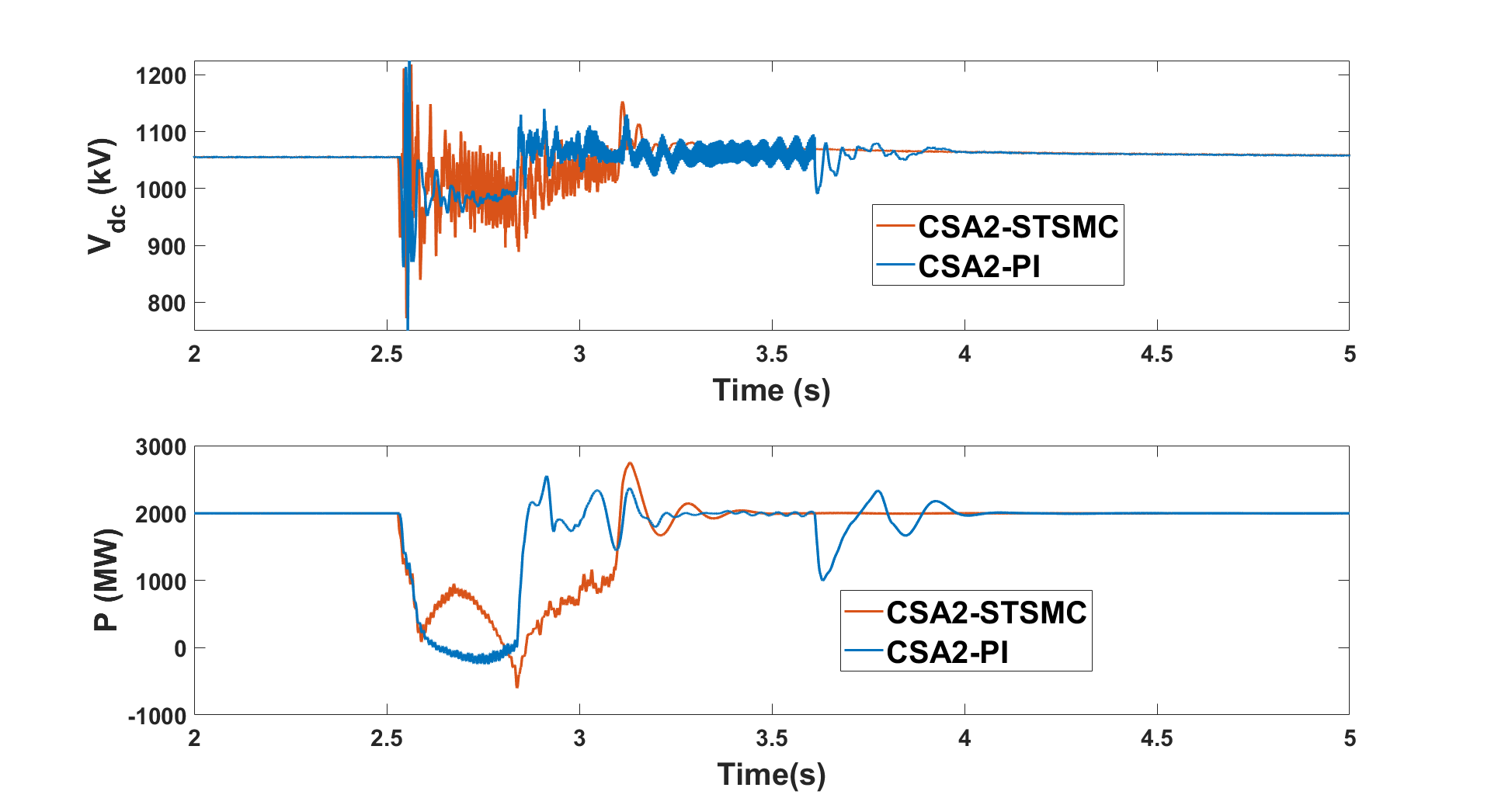}
	\caption{DC link voltage and active power plots for CSA2 converter using PI and STSMC controller during L-L-G faults}
	\label{fig:LLG_PI_STSMC}
\end{figure}

Further, the DC link voltage and active power signals for different faults using STSMC for the inner current loop are given in Fig. \ref{fig:LLG_STSMC}, \ref{fig:LL_STSMC} and \ref{fig:LG_STSMC}. Non-linear controller (STSMC) helps to restore the system back to its stable state quickly irrespective of the type of fault. This is a result of quicker convergence by the sliding surface to the post-fault operating point. A comparison between the PI and STSMC controller when the system is subjected to L-L-G fault is given in Fig. \ref{fig:LLG_PI_STSMC}. As evident, STSMC finds faster settling to the new operating point both for the DC voltage and active power plots, indicating the effectiveness of STSMC controller compared to the state-of-the-art PI controller. 


\section{CONCLUSION}
This paper explores the impact of negative current suppression control during an unbalanced offshore AC fault. The control strategy is applied to a three-terminal meshed MMC-HVDC system that links an offshore wind plant to two onshore AC grids. Various fault scenarios such as line-to-ground, line-to-line, and line-to-line-to-ground faults are analyzed using this control method. The study demonstrates that controlling sequence current components is more effective than voltage amplitude and phase angle control of GFM converters in restoring system stability after fault clearance. The paper also discusses the advantages of using a non-linear controller, particularly focusing on the STSMC controller, over a linear one. The findings indicate that the STSMC controller achieves faster restoration of system stability than the conventional PI controller. Finally for future research, it would be beneficial to identify the optimal level of negative sequence current that ensures system stability while also improving the capabilities of the protection system. 

\bibliographystyle{IEEEtran}
\bibliography{mybib}

\begin{thebibliography}{10}
\providecommand{\url}[1]{#1}
\csname url@samestyle\endcsname
\providecommand{\newblock}{\relax}
\providecommand{\bibinfo}[2]{#2}
\providecommand{\BIBentrySTDinterwordspacing}{\spaceskip=0pt\relax}
\providecommand{\BIBentryALTinterwordstretchfactor}{4}
\providecommand{\BIBentryALTinterwordspacing}{\spaceskip=\fontdimen2\font plus
\BIBentryALTinterwordstretchfactor\fontdimen3\font minus \fontdimen4\font\relax}
\providecommand{\BIBforeignlanguage}[2]{{%
\expandafter\ifx\csname l@#1\endcsname\relax
\typeout{** WARNING: IEEEtran.bst: No hyphenation pattern has been}%
\typeout{** loaded for the language `#1'. Using the pattern for}%
\typeout{** the default language instead.}%
\else
\language=\csname l@#1\endcsname
\fi
#2}}
\providecommand{\BIBdecl}{\relax}
\BIBdecl

\bibitem{ref12}
Y.~Zhang, A.~Shotorbani, L.~Wang, and W.~Li, ``Distributed voltage regulation and automatic power sharing in multi-terminal hvdc grids,'' \emph{IEEE Transactions on Power Systems}, vol.~35, no.~5, 2020.

\bibitem{ref5}
E.~Mehrasa, M.and~Pouresmaeil, S.~Zabihi, and J.~P.~S. Catalão, ``Dynamic model, control and stability analysis of mmc in hvdc transmission systems.'' \emph{IEEE Transactions on Power Delivery}, vol.~32, no.~3, pp. 1471--1482, 2017.

\bibitem{ref4}
J.~Deng, L.~Yao, F.~Cheng, R.~Chen, and X.~Li, ``An enhanced voltage correction control strategy for asymmetrical fault ride-through of offshore wind power using mmc-hvdc.'' \emph{2023 6th International Conference on Energy, Electrical and Power Engineering}, 2023.

\bibitem{roose2021}
T.~Roose, A.~Lekić, M.~M. Alam, and J.~Beerten, ``Stability analysis of high-frequency interactions between a converter and hvdc grid resonances,'' \emph{IEEE Transactions on Power Delivery}, vol.~36, no.~6, pp. 3414--3425, 2021.

\bibitem{ref6}
O.~Göksu, N.~A. Cutululis, P.~Sørensen, and L.~Zeni, ``Asymmetrical fault analysis at the offshore network of hvdc connected wind power plants,'' \emph{2017 IEEE Manchester PowerTech conference proceedings}, 2017.

\bibitem{ref7}
J.~Yan, Q.~Zeng, Y.~Li, and L.~Lin, ``Research on negative sequnence current coordinated control strategy for severe asymmetric offshore ac faults of wind power transmission system based on mmchvdc,'' in \emph{2021 Annual Meeting of CSEE Study Committee of HVDC and Power Electronics (HVDC 2021)}, vol. 2021, 2021, pp. 45--50.

\bibitem{ref8}
M.~Ndreko, M.~Popov, A.~A. van~der Meer, and M.~A. M.~M. vdn~der Meijden, ``The effect of the offshore vsc-hvdc connected wind power plants on the unbalanced faulted behavior of ac transmission systems,'' \emph{IEEE Int. Energy Conference (ENERGYCON)}, 2016.

\bibitem{ref1}
L.~Shi, G.~P. Adam, R.~Li, and L.~Xu, ``Control of offshore mmc during asymmetric offshore ac faults for wind power transmission.'' \emph{IEEE Journal of Emerging and Selected Topics in Power Electronics.}, vol.~8, no.~2, pp. 1074--1083, 2020.

\bibitem{ref2}
H.~Ye, W.~Chen, H.~Wu, W.~Cao, G.~He, and G.~Li, ``Enhanced ac fault ride-through control for mmc-integrated system based on active pcc voltage drop.'' \emph{Journal of Modern Power Systems and Clean Energy}, 2023.

\bibitem{ref10}
R.~Rosso, X.~Wang, M.~Liserre, X.~Lu, and S.~Engelken, ``Grid-forming converters: an overview of control approaches and future trends,'' in \emph{2020 IEEE Energy Conversion Congress and Exposition (ECCE)}, 2020, pp. 4292--4299.

\bibitem{ref13}
L.~Liu, A.~Shetgaonkar, and A.~Lekić, ``Interoperability of classical and advanced controllers in mmc based mtdc power system,'' \emph{International Journal of Electrical Power and Energy Systems}, vol. 148, 2023.

\bibitem{9968871}
M.~Aghahadi, L.~Piegari, A.~Lekić, and A.~Shetgaonkar, ``Sliding mode control of the mmc-based power system,'' in \emph{IECON 2022 – 48th Annual Conference of the IEEE Industrial Electronics Society}, 2022, pp. 1--6.

\bibitem{ref14}
Y.~Jin, Q.~Xiao, Y.~Ji, and T.~Dragicevic, ``A dual-layer back-stepping control method for lyapunov stability in modular multilevel converter based statcom,'' \emph{IEEE Transactions on Industrial Electronics}, vol.~69, no.~3, 2022.

\end{thebibliography}

\end{document}